\documentclass[english]{cimento}
\usepackage{amsmath}
\usepackage{amsfonts}
\usepackage{amssymb}
\usepackage{babel}
\usepackage{graphicx}
\usepackage{scrpage}
\renewcommand\ln{\ell{n}}
\newcommand\beq{\begin{equation}}
\newcommand\eeq{\end{equation}}
\newcommand\bea{\begin{eqnarray}}
\newcommand\eea{\end{eqnarray}}
\newcommand\bseq{\begin{subequations}} %solo con amsmath
\newcommand\eseq{\end{subequations}}
\renewcommand\ln{{\rm ln}}

\title{General Aspects of the de Sitter phase }

\author{Giovanni Imponente \from{inst:federico}
														\from{inst:infn}
														\from{inst:icra}
				\atque Giovanni Montani \from{inst:sapienza}
														\from{inst:icra}}
														
\instlist{\inst{inst:federico} Dipartimento di Fisica, 
 						Universit\'a di Napoli ``Federico II'', Napoli -- Italy 
					\inst{inst:infn} INFN -- Sezione di Napoli
					\inst{inst:icra} ICRA -- International Center for
								Relativistic Astrophysics 
					\inst{inst:sapienza} Dipartimento di Fisica -- G9
					Universit\'a di Roma ``La Sapienza'', Roma -- Italy
					}
					
\PACSes{\PACSit{98.80.Bp }{ Origin and formation of the Universe}
				\PACSit{98.80.Cq }{Inflationary Universe } 
				}					
					
\begin{document}
\maketitle
\begin{abstract}	
 
We present a detailed discussion of the inflationary 
scenario in the context of inhomogeneous cosmologies. 
After a review of the fundamental features characterizing 
the inflationary model, as referred to a homogeneous 
and isotropic Universe, we develop a generalization 
in view of including small inhomogeneous corrections 
in the theory. \\
A second step in our discussion is devoted to show that 
the inflationary scenario provides a valuable dynamical 
``bridge'' between a generic Kasner-like regime and a 
homogeneous and isotropic Universe in the horizon scale.\\
This result is achieved by solving the Hamilton-Jacobi 
equation for a Bianchi IX model in the presence 
of a cosmological space-dependent term. \\
In this respect, we construct a quasi-isotropic 
inflationary solution based on the expansion of the Einstein 
equations up to first-two orders of approximation, 
in which the isotropy of the Universe is due to 
the dominance of the scalar field kinetic term;
the first order of approximation corresponds 
to the inhomogeneous corrections and is driven by 
the matter evolution. \\
We show how such a quasi-isotropic solution contains 
a certain freedom in fixing the space functions 
involved in the problem. 
The main physical issue
of this analysis corresponds to outline the impossibility 
for the classical origin of density perturbations, due to 
the exponential decay of the matter term during the 
de Sitter phase. 
 
\end{abstract}

\section{Introduction}

The homogeneity and isotropy of the Very 
Early Universe evolving  towards the initial 
singularity shows instability for density 
perturbations \cite{LK63}. 

We will discuss how to connect the Mixmaster dynamics
and its properties 
with an inflationary scenario, 
as well as how such a scheme can be interpolated
with a quasi-isotropic 
solution of the Einstein equations 
\cite{IM03, IM03a, IMmg10a}, 
in order to recover the homogeneous 
and isotropic Universe so far observed.

A significant 
degree of inhomogeneity is manifested at 
the scale of galaxies, 
clusters, etc., nevertheless   
on sufficiently large scales 
(of the order of 100 Mpc) inhomogeneities 
are smoothed out and isotropy and 
homogeneity are reached like in the 
Friedmann-Lemaitre-Robertson-Walker 
(FLRW) model.
Even the early Universe exhibits a level of isotropy 
and homogeneity by far higher than now, as
testified by the extreme uniformity 
(of the order of $10^{-4}$) 
of the Cosmic Microwave Background Radiation 
(CMBR).

The homogeneity and isotropy of the Universe
is well tested up to $10^{-3}-10^{-2}$ seconds
of its life by the very good agreement between 
the abundances of light elements as 
predicted by the Standard Cosmological Model 
(SCM) and  the one actually observed \cite{KT90}; 
furthermore, reliable indications support the idea that
the Universe evolved through an 
inflationary scenario and since that age it 
reached isotropy and homogeneity  on the 
horizon scale at least \cite{dB01, map}. 

Despite of these evidences favourable to the large scale 
homogeneity, many relevant 
features suggest that in the very early stage of evolution 
it had to be described by much more general 
inhomogeneous solutions of the Einstein equations.
With respect to this we remark the following two issues:
\begin{enumerate}
\renewcommand{\labelenumi}{({\it\roman{enumi}})}

	\item 	the backward instability of the density perturbations
			to an isotropic and homogeneous Universe \cite{LK63}
			allows us to infer that,
			when approaching the initial singularity, the dynamics 
			evolved as more general and complex models;
			thus the ap\-pea\-rance of an oscillatory regime
			is expected in view of its general nature, 
			\idest  
			because its perturbations correspond simply to 
			redefine the spatial gradients involved in the 
			Cauchy problem.

	\item		Since the Universe dy\-na\-mics during the Planckian 
			era underwent a quantum regi\-me, 
			then no symmetry restrictions can be \textit{a priori}
			imposed on the cosmological model; 
			in fact, the wave functional of the Universe 
			can provide information on the casual scale at most, 
			and therefore the requirement for a global 
			symmetry to hold would imply a large scale 
			correlation of different horizons.
			Thus, the quantum evolution of the Universe is 
			appropriately described only in terms of a generic 
			inhomogeneous cosmological model.

	\end{enumerate}

%			the spectrum of the Cosmic Microwave Background (CMB)
%			has to increase in amplitude 

In this paper, we address two main topics concerning 
the Universe evolution: in first place, we show how 
the inflationary scenario provides the natural 
framework to link the inhomogeneous oscillatory 
regime with anisotropic and homogeneous Universe on the 
horizon scale; then, we investigate how  
a classical origin of inhomogeneous perturbations
is ruled out out by a de Sitter phase in a quasi-isotropic 
scheme.

\section{The Standard Cosmological Model}

%\subsection{The Robertson-Walker Metric}

The most ge\-ne\-ral me\-tric which is spa\-tial\-ly 
homo\-ge\-neous and iso\-tropic is the FLRW one, which in terms of the 
\textit{comoving coordinates} 
$(t, r, \theta, \phi)$ reads
\begin{equation}
\label{lfried}
ds^2=dt^2-R^2(t)\left(\frac{dr^2}{1-kr^2}+
r^2d{\theta}^2+r^2\sin^2{\theta}d{\phi}^2\right)
\end{equation}
where the scale factor  $R(t)$ is a generic 
function of time only and, for an appropriate 
rescaling of the coordinates, the factor 
$k=0,\pm1$  distinguishes the sign  
of constant spatial curvature. \\
Such coordinates represent a reference frame 
participating in the expansion of the Universe: 
an observer at rest  will remain at rest,
leaving the effects of the expansion to the 
cosmic scale factor $R(t)$.

%\subsection{The Friedmann Equation}

The FLRW dynamics is reduced to the time 
dependence of the scale factor $R(t)$ once solved
 the Einstein equations %(\ref{fieex})
with a stress-energy tensor $T_{\mu\nu}$ 
for all the fields present (matter, radiation, etc.)
which then must be diagonal, with the spatial 
components equal to each other for the homogeneity and 
isotropy constraints. 
A simple realization is given by the  
perfect fluid one, characterized by a space-independent
energy density $\rho(t)$ and pressure $p(t)$
	\begin{equation}
	T^\mu{}_\nu=\textrm{diag}(\rho,-p,-p,-p) \, ,
	\end{equation}
and in this case the $0-0$ component of the Einstein 
equations reads 
\begin{equation}
\label{frieq1}
\frac{{\dot R}^2}{R^2}+\frac{k}{R^2}=\frac{8\pi G}{3}\rho
\end{equation}
which is the {\it Friedmann equation},
while the $i-i$ components are 
\begin{equation}
\label{frieq2}
2\frac{\ddot R}{R}+\frac{{\dot R}^2}{R^2}+
\frac{k}{R^2}=-8\pi Gp \, .
\end{equation}
The difference between (\ref{frieq1}) and 
(\ref{frieq2}) leads to
	\begin{equation}
	\frac{\ddot R}{R}=-\frac{4\pi G}{3}(\rho+3p)	\, ,
	\end{equation}
which is solved  for  $R(t)$ once provided an 
equation of state, \idest a relation between  
$\rho$ and $p$. \\
When the Universe was radiation-dominated, 
as in the early period, 
the radiation component provided the greatest 
contribution to its energy density
and for a photon gas we have $p=\rho/3$. \\
The present-time Universe  is, on the
contrary, matter-dominated: the ``particles'' 
(\idest the galaxies) 
have only long-range (gravitational) interactions
and can be treated as a pressureless gas (``dust''):
the equation of state is $p=0$.

Nevertheless, the energy tensor appearing in the 
Einstein field equations
describes the complete local energy due to all 
non-gravitational fields, while gravitational
energy has a non-local contribution. 
An unambiguous formulation for such a non-local 
expression is found only in the expressions
used at infinity for an asymptotically flat
space-time \cite{PEN82, ADM}. 
This is due to the property of the mass-energy 
term to be only one component of the 
energy-momentum tensor which can be reduced
only in a peculiar case
to a four-vector expression which can not 
be summed in a natural way. 

Bearing in mind such difficulties,
the conservation law 
$T^{\mu\nu}{}_{;\nu}=0$  leads to
	\begin{equation}
	d(\rho R^3)=-pd(R^3)
	\end{equation}
\idest the first law of thermodynamics in an 
expanding Universe which, via the equation 
of state, leads to a differential relation for 
$\rho(R)$. 
If we take its solution together 
with the Friedmann equation (\ref{frieq1}) 
corresponding to $k=0$, then we 
find the following behaviours
\begin{subequations}
\begin{eqnarray}
\textrm{RADIATION\qquad} 
&\rho& \propto R^{-4}\, , \quad R \propto t^{1/2}  \\
	\textrm{MATTER \qquad} 
	&\rho& \propto R^{-3}\, , \quad R \propto t^{2/3}
\end{eqnarray}
\end{subequations}
where as long as the Universe is not
curvature-dominated, \idest for sufficiently 
small values of $R$, 
the choice of $k=0$ is not relevant.
The equation of state $p=-\rho$
leads to 
$\rho=\textrm{const.}$ and  $R\propto e^{At}$ 
$A=const.$,
\idest a phase of exponential expansion, equivalent 
to adding a constant term to the right-hand side 
of the Einstein equation mimicking a cosmological 
constant: this is exactly what the 
inflationary paradigm proposes to overcome the 
paradoxes of the standard model outlined in the 
next Section \ref{paradoxs}. \\
The Hubble parameter $H\equiv {\dot R}/R$ and 
the critical density ${\rho}_c \equiv 3H^2/8\pi G$
make it possible to rewrite (\ref{frieq1}) as
\begin{equation}
\label{fried3}
\frac{k}{H^2R^2}=\frac{\rho}{3H^2/8\pi G}-1\equiv\Omega-1,
\end{equation}
where $\Omega$ is the ratio of the density 
to the critical one $\Omega\equiv\rho/{\rho}_c$;
since $H^2R^2$ is always positive, the relation 
between the sign of $k$ and the sign of $(\Omega-1)$ 
reads
	\begin{subequations}
	\begin{eqnarray}
	k=+1\quad &\Rightarrow &\quad\Omega>1 
												\qquad\textrm{CLOSED} \\
	k=0\quad &\Rightarrow &\quad\Omega=1 \qquad\textrm{FLAT}\\
	k=-1\quad &\Rightarrow &\quad\Omega<1 \qquad\textrm{OPEN} %\nonumber
	\end{eqnarray}
	\end{subequations}

\section{Shortcomings of the Standard Model: 
Horizon and Flatness Paradoxes\label{paradoxs}}

Despite the simplicity of the Friedmann solution 
(in view also of the thermodynamic property which can 
be studied in detail) some paradoxes occur when 
taking into account the problem
of \textit{initial conditions}.
The observed Universe has to match very specific 
physical conditions in the very early epoch, but 
\cite{CH73} %Collins and Hawking 
 showed that 
the set of initial data that can evolve to a 
Universe similar to the present one is of zero measure
and the standard model tells  {\it nothing} 
about initial conditions.

\subsection*{Flatness}

Let us assume  that all particle species present 
in the early Universe have
the same temperature as the photon bath, 
\idest  
$T_i=T_\gamma$ and are far from
any mass thre\-shold. 
Then the average number of degrees of freedom 
for the photons and fermions bath $g^*$ is 
a constant and 
$T\propto R^{-1}$ \cite{KT90}. 
The average energy density 
\beq
\rho = \frac{\pi^2}{30} g^* T^4
\eeq
substituted in the (\ref{frieq1}) reads 
	\begin{equation}
	\label{mlpp}
	\left( \frac{\dot T}T \right)^2 + \epsilon(\,T)T^2=\frac{4\pi^3}{45}Gg^*T^4 \, ,
	\end{equation}
where
	\begin{equation}
	\label{mlep}
	\epsilon(\,T)\equiv\frac k{R^2T^2}=k\left[ \frac{2\pi^2}{45}\frac{g^*}S \right]^{2/3}  \, , 
	\end{equation}
and since $S=R^3s$ 	is the total entropy 
per comoving volume 
the entropy density reads 
\beq
s=\frac{2\pi^2}{45} g^* T^3 \, .
\eeq
Today $\rho\simeq\rho_c$, then by taking 
$\rho<10\rho_c$ in
(\ref{fried3}) we have
	\begin{equation}
	\left|\frac k{R^2} \right|<9H^2 \, .
	\end{equation}
For $k=\pm 1$ (the case $k=0$ is regained 
in the limit $R\to\infty$), we have for today
$R>\frac13H^{-1}\approx3\cdot10^9$ years
and $T_{\gamma}\simeq$ 2.7 K; 
 from (\ref{mlep}), the present photon contribution 
 to the entropy  has the lower bound
	\begin{equation}
	S_\gamma>3\cdot10^{85}
	\end{equation}
expressed in units 
of the Boltzmann constant
$k_B=1.3806\cdot10^{-16}$ erg/K.
The relativistic particles present today together with 
photons are the three neutrino species and 
their contribution to the total entropy 
is of the same order of magnitude
	\begin{equation}
	S>10^{86} \, ,
	\end{equation}
and finally with 
	\begin{equation}
	|\epsilon|<10^{-58}{g^*}^{2/3}
	\end{equation}
we gain 
	\begin{equation}
	\label{drrh}
	\left| \frac{\rho-\rho_c}{\rho} \right|=\frac{45}{4\pi^3} \frac{m_P^2}{g^*T^2}|\epsilon|
	<3\cdot10^{59}{g^*}^{-1/3} \left( \frac{m_P}T \right)^2 \, ,
	\end{equation}
where $m_P$ is the Planck mass 
$\displaystyle \left( {\hbar c}/{G} \right)^{1/2}=
2.1768\cdot10^{-5}$g 
$=1.2211\cdot~10^{19}$ GeV. \\
When taking $T=10^{17}$ GeV, all species in 
the standard model of particle interactions 
-- 8 gluons, W$^\pm$, Z$^0$, 3 generations
of quark and leptons -- are relevant and 
 relativistic: then $g^*\approx 100$ 
and finally 
\begin{equation}
	\left| \frac{\rho-\rho_c}{\rho} \right|_{T=10^{17}
	\mathrm{GeV}}<10^{-55} \, .
	\end{equation}
A flat Universe today requires $\Omega$ 
of the original one close to 
unity up to a part in $10^{55}$. 
A little displacement from flatness at the beginning
-- for example $10^{-30}$ --
 would produce an actual Universe either very 
 open or very closed, so that $\Omega=1$ 
 is a very unstable condition: this is 
 the \textit{flatness problem}. \\
The natural time scale for cosmology is 
the Planck time
($\sim 10^{-43}$ sec): in a time of this order
a typical closed Universe would  reach 
the maximum size while an open one would become curvature 
dominated.
The actual Universe has survived $10^{60}$ 
Planck times without neither recollapsing nor becoming 
curvature dominated.

\subsection*{Horizon}

Neglecting the $\epsilon T^2$ term,
(\ref{mlpp}) is solved by
\begin{equation}
T^2= \left(  \frac{4\pi^3}{45} g^* \right)^{-1/2} 
\frac{m_P}{2\,t} \, .
\end{equation}
A light signal emitted at $t=0$ travelled 
during a  
time $t$ the physical distance 
	\begin{equation}
	\label{ldist}
	l(t)=R(t)\int_0^{\,t}\frac{dt'}{R(t')}=2\,t
	\end{equation}
in a radiation-dominated Universe with 
$R\propto t^{1/2}$, measuring  
the physical horizon size, \idest the linear 
size of the greatest region causally connected 
at time $t$.
The  distance (\ref{ldist}) has to be compared with the 
radius $L(t)$ of the region which will evolve
in our currently observed region of the Universe. 
Conservation of entropy for $s\propto T^3$
gives
	\begin{equation}
	\label{ellel}
	L(t)=\left( \frac{s_0}{s(t)} \right) ^{1/3}L_0 \, ,
	\end{equation}
where $s_{0}$ is the present entropy density 
and $L_0\sim H^{-1}\simeq10^{10}$ years is 
the radius of the currently observed region of 
the Universe. The ratio of the volumes provides
	\begin{equation}
	\label{elle3}
	\frac{l^3}{L^3}=4\cdot10^{-89} g*^{-1/2} \left( \frac{m_P}{T} \right)^3
	\end{equation}
and, as above, for $g^*\sim 100$ and 
$T\sim 10^{17}\mathrm{GeV}$ we obtain 
	\begin{equation}
	\left. \frac{l^3}{L^3} \right|_{T=10^{17}\mathrm{GeV}} 
	\sim 10^{-83} \, .
	\end{equation}
The currently observable Universe is composed 
of several regions which have \textit{not} been in 
causal contact for the most part of their evolution,
preventing an explanation
about the present days Universe smoothness. 
In particular, the spectrum of the CMBR 
is uniform up to  $10^{-4}$. 
Moreover, we have at the time of 
recombination, \idest when the photons of the CMBR last 
scattered, the ratio $l^3/L^3\sim 10^5$: 
the present Hubble volume consists of about 
$10^5$ causally disconnected
regions at recombination and no process could 
have smoothed out the temperature differences 
between these regions without violating causality. 
The particle horizon at recombination
subtends an angle of only $0.8^\circ$ in the sky 
today, while the CMBR is uniform across the sky.

\section{The Inflationary Paradigm}

The basic ideas  of the 
theory of inflation rely firstly on the 
original work by  \cite{G81}, %Guth, 
\idest the \textit{old inflation}, 
which provides a phase in the Universe evolution 
of exponential expansion; then  
 %Linde's 
 formulation of \textit{new inflation} 
 by \cite{Linde1} introduced the  
 slow-rolling phase in inflationary dynamics; 
finally,  many models  have sprung from 
the original theory.

%\subsection{Old Inflation: the Original Idea}

In   \cite{G81} % 1981 Alan H. Guth 
 is described a scenario  capable
of avoiding the horizon and flatness problems:
 both paradoxes would disappear 
dropping  the assumption of adiabaticity and in 
such a case the entropy per comoving volume $s$
would be related as
	\begin{equation}
	\label{entro}
	s_0=Z^3 s_{\mathrm{early}}
	\end{equation}
where $s_0$ and $s_{\mathrm{early}}$ refer
to the values at present and at very early times, 
for example at $T=T_0=10^{17}~\mathrm{GeV}$, 
and Z is some large factor. \\
With this in mind, the right-hand side of (\ref{drrh}) is 
multiplied by a factor $Z^2$ and the value
of $|\rho-\rho_c|/\rho$ would be of the  order of 
unity if 
	\begin{equation}
	Z>3\cdot 10^{27} \, ,
	\end{equation}
getting rid of the flatness problem.\\
The right-hand side of (\ref{ellel}) is multiplied by
$Z^{-1}$: for  any
given temperature, the length scale 
of the early Universe is smaller by a factor $Z$ 
than previously evaluated, and for $Z$ 
sufficiently large the initial region which 
has evolved in our observed one would have 
been smaller than the horizon size at that time. \\
Let us  evaluate $Z$ considering that the 
right-hand side of (\ref{ellel})
is multiplied by $Z^3$: if
	\begin{equation}
	Z>5\cdot 10^{27} \, ,
	\end{equation}
then the horizon problem disappears. \\
Making some \textit{ad hoc}
assumptions, the model accounts for the 
horizon and flatness paradoxes while a
suitable theory needs a physical process 
capable of such a large entropy production.
A simple solution relies on the assumption 
that at very early times  the energy
density of the Universe was dominated by 
a scalar field $\phi(\vec{x}, t)$,
\idest 
$\rho=\rho_{\phi}+\rho_{\mathrm{rad}}+
\rho_{\mathrm{mat}}+\dots$ with
$\rho_\phi\gg\rho_{\mathrm{rad,\;mat,\;etc}}$
and hence $\rho\simeq\rho_{\phi}$.

The quantum field theory  Lagrangian density 
for such a field is 
	\begin{equation}
	{\cal L}=\frac{\partial^{\mu}\phi\,
	\partial_{\mu}\phi}2-V(\phi)
	\end{equation}
and the corresponding stress-energy tensor 
	\begin{equation}
	T^{\mu}_{~\nu}=\partial^{\mu}\phi\,
	\partial_{\nu}\phi-{\cal L}\delta^{\mu}_{~\nu}
	\end{equation}
by which for a spatially homogeneous and isotropic
Universe the form  of a perfect fluid
leads to \cite{pow}
\bseq
\begin{eqnarray}	
\label{rphi}
	\rho_\phi&=&\frac{\dot{\phi}}{2}^2+ V(\phi)+
	\frac{1}{2R^2} \nabla^2\phi	\, , \\
	\label{pphi}
	p_\phi&=&\frac{\dot{\phi}}{2}^2-V(\phi)-
	\frac{1}{6R^2}\nabla^2\phi \, .
	\end{eqnarray}
	\label{rpphi}
\eseq
Spatial homogeneity would induce a slow 
variation of $\phi$ with position, hence
the spatial gradients are negligible and 
the ratio $\omega=p/\rho$ reads
\begin{equation}
\label{psur}
\frac{p_\phi}{\rho_\phi}\simeq	\frac{\displaystyle\frac{\dot{\phi}}{2}^2+V(\phi)}
{\displaystyle\frac{\dot{\phi}}{2}^2-V(\phi)} \, .
	\end{equation}
If the field is at a minimum of 
the potential, $\dot{\phi}=0$, and 
(\ref{psur}) becomes an equation of state 
\begin{equation}
\label{eqst}
	p_\phi=-\rho_\phi
	\end{equation}
giving rise to a phase of exponential growth 
of $R\propto e^{Ht}$,  the {\it inflationary} 
or \textit{de Sitter} {\it phase}.\\
The field evolution is very different when 
in vacuum or in a thermal bath and  
such a coupling can be summarized 
by adding a term $-(1/2)\lambda T^2\phi^2$ to 
the Lagrangian.   
The potential $V(\phi)$ is replaced by 
the \textit{finite-temperature} effective 
potential
	\begin{equation}
	\label{potphi}
	V_T(\phi)=V(\phi)+\frac12\lambda T^2\phi^2 \, .
	\end{equation}
In the old inflation, $V(\phi)$ appearing  in (\ref{potphi})
has the form of a Georgi-Glashow $SU(5)$ theory
	\begin{equation}
	\label{georgi}
	V(\phi)= \frac{1}{4} \phi^4 - \frac{1}{3} (A +B ) \phi^3 +
	\frac12AB\phi^2
	\end{equation}
with 
%$(1/2)A>B>0$, 
$A>2B>0$, and possesses   a local minimum 
at $\phi=0$ and a global minimum at $\phi=A$, 
separated by a barrier with a maximum at $\phi=B$. 
%\begin{figure}[t]
%\begin{center}
%\includegraphics{Vphi2}
%\caption[``Bare'' Georgi-Glashow potential]
%{\small Qualitative behaviour of the ``bare'' potential (51). 
%We see it has a local minimum at
%$\phi=0$ and a global minimum at $\phi=A$, 
%separated by a barrier at $\phi=B$.}
%\end{center}
%\end{figure}
The temperature-dependent term 
$-(1/2)\lambda T^2\phi^2$ leaves
the local minimum unchanged and raises the 
global minimum as well as the maximum -- 
the former by a larger amount than the latter. \\
At sufficiently high temperature $V_T(\phi)$ 
has only one global minimum at $\phi=0$ 
and, as long as  
$T$ decreases,  a second
minimum develops at  $\phi=\sigma(T)$,
with $V(0)<V(\sigma)$, and $\phi=0$ is still the 
true minimum of the potential. \\
At a certain critical temperature $T_c$ the two 
minima are exactly
degenerate as $V(0)=V(\sigma)$ and at temperatures 
below $T_c$,  $V(\sigma)<V(0)$ and
$\phi=0$ is no longer the true minimum of 
the potential. \\

Let us  consider when  
at some initial time, corresponding to $T=T_i>T_c$, 
the field $\phi$ is trapped in the minimum at 
$\phi=0$ (\textit{false vacuum}) with
constant energy density, given by (\ref{rphi}) 
$V_T(0) \simeq T_C^4$. The temperature lowers
with the Universe expansion up to
the critical value $T_c$: 
the scalar field begins dominating the Universe and 
a second minimum of the potential develops 
at $\phi=\sigma$.
The  inflationary phase is characterized by 
	\begin{equation}
	\label{infrexp}
	R(t)\propto e^{Ht}
	\end{equation}
where the Hubble parameter $H\equiv\dot R/R$ is 
a constant;  $\phi=0$ becomes a metastable 
state, since there exists a more 
energetically favourable one 
at $\phi=\sigma$ (\textit{true vacuum}). \\
The potential barrier lying within cannot
prevent a non-vanishing probability per unit 
time that the 
field performs  a first order phase transition 
via quantum tunnelling to the true vacuum state,
proceeding along by bubble nucleation: bubbles 
of the true vacuum phase are 
created expanding outward at the speed of light 
in the surrounding ``sea'' of false vacuum,
until all the Universe has undergone the phase 
transition. \\
If  the rate of bubble nucleation is low, 
the time  to
complete the phase transition can be very long 
if compared to the expansion time scale:
when the transition ends, the Universe has cooled 
to a temperature $T_f$ many orders of magnitude 
lower than $T_c$. \\
On the contrary, when  approaching the true vacuum, 
the field $\phi$ begins to oscillate around this 
position on a time scale short 
if compared to the expansion one, releasing all 
its vacu\-um energy in the form of $\phi$-particles, 
the quanta of the $\phi$ field. 
The oscillations are 
damped by particles decay and, when the 
corresponding products 
thermalize, the Universe is
reheated to a temperature $T_r$ of the order of $T_c$. 
This represents the release of 
the latent heat associated with the phase transition
after which the scalar field is no longer
the dominant component of the Universe: 
inflation comes to an end and the standard
FLRW cosmology is recovered. \\

The non-adiabatic reheating process
releases an enormous amount of entropy whose
density is increased by a factor
of the order of $(T_r/T_f)^3\simeq (T_c/T_f)^3$, 
while $R$ remains nearly constant and the entropy 
is increased by a factor $(T_c/T_f)^3$ too. \\
Flatness and horizon paradoxes are solved 
if the Universe super-cools of 28 or more orders 
of magnitude during inflation. Even if this looks very
difficult to achieve, it is 
enough that the transition takes place in 
about a hundred Hubble times: the inflationary 
expansion is adiabatic and the entropy 
density $s\sim R^{-3}$. Since $s\propto T^3$, 
then $T\propto R^{-1}\propto e^{-Ht}$ and finally
	\begin{equation}
	\label{temph}
	\frac{T_c}{T_f}=e^{H\Delta t} \, ,
	\end{equation}
where $\Delta t$ is the duration of the 
de Sitter phase. From the requirement 
$(T_c/T_f)>10^{28}$ it follows
	\begin{equation}
	\Delta t>50H^{-1} \, .
	\end{equation}
The critical temperature is estimated to be of 
order of the energy typically
involved for Grand Unification Theory spontaneous symmetry-breaking 
phase transition, $10^{14}$ GeV, and 
	\begin{equation}
	H^2=\frac{8\pi G}3\rho=\frac{8\pi G}3V(0)
	\simeq \frac{T_c^4}{m_{P}^2} 
	\qquad \rightarrow \qquad 
	H^{-1}\simeq 10^{-34} \textrm{s} \, .
	\end{equation}
The inflation removes the discussed paradoxes
if the transition takes place in a time 
$t\approx 10^{-32}$s, nevertheless leaving 
some open problems
regarding its dynamics:

\begin{enumerate}

\renewcommand{\labelenumi}{{- \it\alph{enumi} -}}
	\item 	in the old scenario inflation never
			ends, due to smallness of  the tunnelling
			transition rate, so that the nucleation of true 
			vacuum bubbles is rare; 

	\item 		the energy released
		 	during the reheating is stored in the 
			bubbles kinetic
			energy  so that the reheating proceeds via bubble 
			collisions which remain too rare due to low
			transition rate  to produce sufficient reheating:
			the phase transition is never completed;
	
	\item  	such a discontinuous process of bubble nucleation
			via quantum tunnelling should produce a lot of 
			inhomogeneities which aren't actually observed.

\end{enumerate}

\subsection{New Inflation: the Slow 
Rolling Model\label{slowrolling}}

In 1982, both \cite{Linde1} %Linde 
 and 
%Albrecht \& Steinhardt 
\cite{Alb} proposed a variant 
of Guth's model, now referred to
as \textit{new inflation} or 
\textit{slow-rolling inflation}, in order to
avoid the shortcomings
of the old inflation. 
Their original idea  considered 
a different mechanism of symmetry breaking, 
the so-called Coleman-Weinberg (CW) one. 
The potential of the CW model for a gauge boson
field with a
vanishing mass reads as
\begin{equation}
\label{potcw}	V(\phi)=\frac{B\sigma^4}2+B\phi^4
\left[\ln\left(\frac{\phi^2}{\sigma^2}
\right)-\frac12 \right] \, ,
\end{equation}
where $B$ is connected to the fundamental constants 
of the theory and is $\simeq 10^{-3}$, while
$\sigma$ gives the energy associated with the 
symmetry breaking process and is 
$\simeq 2\cdot10^{15}$ GeV. \\
The finite-temperature effective
potential is obtained as above in (\ref{potphi})
by adding a term of 
the form $(1/2)\lambda T^2\phi^2$.  
Expression (\ref{potcw}) can be generalized
by adding a mass term of
the form $-(1/2)m^2\phi^2$. 
Defining  a temperature-dependent ``mass''
	\begin{equation}
	m_T\equiv\sqrt{-m^2+\lambda T^2} \, ,
	\end{equation}
the temperature-dependent potential becomes
\begin{equation}
		V_T(\phi)=\frac{B\sigma^4}2+
		B\phi^4\left[\ln\left(\frac{\phi^2}{\sigma^2}\right)
		-\frac12 \right]
			+\frac12 m^2_T\phi^2
	\label{CW} \, .
	\end{equation}
The quantity ${m_T}^2$ can be used to 
parametrize the potential (\ref{CW}):

	\begin{enumerate}
	\item 	when ${m_T}^2>0$, the point $\phi=0$ 
			is a minimum of the 	potential, while when 
			$m_T^2<0$ it is a maximum;
		
	\item 	when ${m_T}^2 < \displaystyle 
		\frac{4\sigma^2}{e}\simeq 1.5\sigma^2$, 
		a second minimum develops for 
		some $\bar{\phi}>0$; initially 
		this minimum is higher than the one at 0, 
		but when $m_T$ becomes lower than
		a certain value $m_T^*$ $(0<m_T^*<1.5\sigma^2)$ 
		it will eventually become the global
	minimum of the potential.
	\end{enumerate}

If at some initial time the $\phi$-field is 
trapped in the minimum at $\phi=0$ 
the true minimum of 
the potential can eventually disappear
as the temperature lowers. 
In this case, as $m_T$ approaches 0, the 
potential barrier becomes low and can 
 be easily overcome by \textit{thermal} 
(not quantum) tunnelling, 
\idest due to classical (thermal) fluctuations of 
the $\phi$ field around its minimum; 
the barrier can disappear completely when $m_T=0$. 
Independently of what really happens, 
the phase transition doesn't proceed via a
quantum tunnelling -- a very
discontinuous and a strongly first order 
process -- but it evolves either by a 
\textit{weakly} first order 
(thermal tunnelling) or second order 
(barrier disappearing at $m_T=0$). \\
Hence the transition occurs rather smoothly,
avoiding the formation of undesired 
inhomogeneities: inflation is not yet
started, so the requirement
for the field to take a long time to escape 
the false vacuum  is not necessary; 
the transition rate can  be very close to unity, 
and completed without problem. \\
When the $\phi$-field has passed the barrier
(if any), it begins to evolve towards its 
true minimum. 
However, the  potential (\ref{CW})
has a very interesting feature: if the 
coefficient of the logarithmic term is 
sufficiently 
high, the potential is very flat around 0, 
and then the field $\phi$ ``slow rolls'' 
rather than falling
abruptly in the true vacuum state: during this 
slow roll phase the inflation takes place,
lasting enough to produce the required 
supercooling, as seen in the
previous Section. When the field reaches the 
minimum, it begins to oscillate around it thus
originating the reheating. \\
The problems of Guth's originary model 
are skipped moving 
the inflationary phase \textit{after} the field
has escaped the false vacuum state, 
by adding the slow-rolling phase. \\
Virtually all models of 
inflation are based upon this principle.

\section{The Bridge Solution\label{bridgesol}}

An inflationary scenario is of crucial importance
to understand how an anisotropic Universe like 
the one described by the Bianchi type IX cosmological 
solution can approach an isotropic Universe when 
the volume expands enough. \\
In fact, during the inflation, the dominant term 
in the Einstein equations corresponds to the effective
cosmological constant associated with the false-vacuum
energy; such a term is an isotropic one and, 
when it dominates, it produces an exponential decay of 
the Universe anisotropies. \\
In this Section 
we show how it is possible to 
interpolate a Kasner-like behaviour with 
an isotropic stage of evolution. 
We will 
refer to this scheme as the \textit{bridge solution}
as in \cite{KM02}, 
because it allows to match the chaotic dynamics of the 
Bianchi IX model together with the later isotropic 
dynamics of the SCM. 

With respect to this, let us observe that 
in the presence of an effective cosmological 
constant the action of the Bianchi IX framework
 takes the form 
\beq
\delta I = \delta \int \left( p_{\alpha} {\alpha}^{\prime} 
+ p_+ {\beta_+}^{\prime} +p_- {\beta_-}^{\prime}
+ p_{\phi} \phi^{\prime}
 - \sqrt{\frac{3\pi}{2}} N e^{-3\alpha}
 {\cal H} \right) d\eta=0
\label{briact}
\eeq
where we recall that  ${\cal H}$ reads as
\beq
2 {\cal H} = -p^2_{\alpha} +p^2_+ +p^2_- + p^2_{\phi}
+{\cal V}  + e^{6 \alpha}\Lambda \, ,
\label{brih}
\eeq
$\Lambda$ denotes  a constant term
and the Bianchi IX potential is  \cite{MTW}
\beq
{\cal V}=\frac{e^{4\alpha}}{3}
\Big\{  e^{-8 \beta_+} 
-4 e^{-2 \beta_+}\cosh(2\sqrt{3}\beta_-)
+2 e^{4 \beta_+}
\big[ \cosh(4\sqrt{3}\beta_-) -1\big]\Big\} \, .
\label{bripot}
\eeq
The variation of the action (\ref{briact}) with 
respect to $N$ provides the super-hamiltonian
constraint to be ${\cal H} =0$. \\
Near the Big Bang, $\alpha \rightarrow - \infty$,
the Bianchi IX potential (\ref{bripot}) turns out to be 
negligible with respect to the cosmological 
constant term and then, by replacing the conjugate 
momenta as 
\beq
p_X \rightarrow \frac{\partial I}{\partial X} \, ,
\qquad X= \alpha, \beta_{\pm}, \phi
\eeq
we get the Hamilton-Jacobi equation
\beq
 -\left(\frac{\partial I}{\partial \alpha}\right)^2
+ \left(\frac{\partial I}{\partial \beta_+}\right)^2   
+ \left(\frac{\partial I}{\partial \beta_-}\right)^2   
+ \left(\frac{\partial I}{\partial \phi}\right)^2   
+ e^{3 \alpha}\Lambda =0 \, .
\label{brihj}
\eeq
The general solution of (\ref{brihj})
takes the form 
\beq
I({\chi }^{r}, \alpha )=  \sum_{r}K_{r} {\chi}^{r}
+\frac{2}{3}K_{\alpha }+{\frac{K}{3}} 
\ln \left| {\frac{K_{\alpha }-K}{K_{\alpha }+K}}\right| \, ,
\label{brihjsol}
\eeq
where $\chi_r = \{\beta_+, \beta_-, \phi\}$, 
$K_r$ are constants of integration and 
\beq
K= \sqrt{\sum_{r}K_{r}^{2}}
\eeq
in which the index $r$ is the label for 
$\beta_{\pm}, \phi$, 
while 
\beq
K_{\alpha }(K_r, \alpha )
=\pm \sqrt{ \sum_{r}K_{r}^{2}+6 \Lambda 
\exp \left( 3\alpha \right) } \, ,
\eeq
by which we adopt the negative sign in order to 
describe Universe expansion; 
in fact, we have 
\beq
\frac{\partial \alpha}{\partial t} =
- \sqrt{\frac{3\pi }{2}}N e^{-3 \alpha} p_{\alpha} \,.
\eeq
According to the Hamilton-Jacobi method, 
we differentiate the action with respect to the quantities
$K^r$ and then, by putting the resulting expressions
equal to arbitrary constant functions as
\beq
\frac{\delta I}{\delta K^{r}}=
{\chi}_{0}^{r}=\textrm{const.} \, ,
\eeq
we find the solutions describing 
the trajectories of the system to be 
\beq
{\chi }^{r}(\alpha)={\chi }_{0}^{r}
+{\frac{K_{r}}{3\left| K\right| }}
\ln \left| {\frac{K_{\alpha }-K}
{K_{\alpha }+K}}\right| \, .
\eeq

Let us now consider the two opposite limits:

\begin{itemize}
\renewcommand{\labelitemi}{ $\alpha \rightarrow - \infty$	}
%	\item 
%	\renewcommand{\labelitemi}{ $\alpha \rightarrow + \infty$	}
%	\item qualcos'altro

%\begin{enumerate}
%\renewcommand{\labelenumi}{{\it\roman{enumi}}.	}
	\item %for $\alpha \rightarrow - \infty$
		 			we find the solution 
		 	 \beq
		 			{\chi}^{r}(\alpha)={\chi}_{0}^{r}
					-{\frac{K_{r}}{K}}(\alpha -\alpha _{0})
			\label{brikas}
			\eeq
		corresponding to a Kasner-like behaviour 
		which can be regarded as the \textit{last
		epoch} of the oscillatory regime 
		\cite{BKL70, IM01};
	
\renewcommand{\labelitemi}{ $\alpha \rightarrow + \infty$	}	
	
	\item %for $\alpha \rightarrow + \infty$
		we get the isotropic stage of evolution 
		\beq
		{\chi }^{r}(\alpha) \rightarrow {\chi}_{0}^{r} \, .
		\label{briiso}
		\eeq
	In fact, when the anisotropies $\beta_{\pm}$ approach 
	constant values, they are no longer dynamical degrees
	of freedom and the solution looks homogeneous and 
	isotropic. In the same limit, the scalar field
	frozens to a constant value too and 
	it disappears from the 
	dynamics as soon as the inflation starts.	
	
\end{itemize}
%\end{enumerate}

Our analysis provides an interpolation between the two 
relevant stages  of the Universe evolution and is a 
convincing feature in favour of the idea that inflation 
can be a mechanism to gain isotropy. 
In this sense, the inflationary scenario constitutes
the natural mechanism by which the chaotic 
dynamics of the Bianchi IX model can be smoothed out
towards a closed FLRW dynamics.

As shown in \cite{KM02} the results of this Section 
can be extended to the generic inhomogeneous solution, 
\idest point by point in space the same 
behaviour above outlined takes place. 

From a cosmological point of view this means 
that after inflation 
the Universe reaches an high degree of 
homogeneity and isotropy 
within each horizon size.

\section{Quasi-isotropic Cosmological Solution
\label{quasiisoKL}}

The perturbations to matter
distribution  not affecting uniformity 
are damped with time or remain constant 
in the isotropic model \cite{LK63}. 
Hence, the evolution backwards in time of small 
density perturbations is of particular interest 
when considering cosmological models more general 
then the homogeneous and isotropic one, the 
assumption of uniformity  being justified only at 
an approximate level.

The Friedmann solution is a particular case
belonging to the
class in which space contracts
in a quasi-isotropic way, in the sense that 
linear distances change with the same 
time-dependence in all directions and
such a solution existing only in a space filled
with matter.

When considered the isotropic solution in the 
synchronous reference frame, isotropy and homogeneity
are reflected in the vanishing of the off-diagonal 
metric components $g_{0\alpha}$. 
The approach to zero of such functions depends 
upon the matter equation of state: 
for the ultra-relativistic 
equation $p=\epsilon/3$, the metric is linear in $t$, 
hence $g_{\alpha \beta }$ is supposed to be
expandable in integer powers of $t$.\\
Corresponding to these assumptions, 
the Einstein equations 
reduce to the partial differential system
\bseq
\begin{align}% \displaystyle
\frac{1}{2} \partial _t k_{\alpha }^{\alpha } 
	+ \frac{1}{4} k_{\alpha }^{\beta }k_{\beta }^{\alpha } 
	&= 	\frac{\epsilon }{3} \left(4 u_0 u^0  + 1\right) \\  
\frac{1}{2}\left( k^{\beta }_{\alpha ;\beta } - 
	k^{\beta }_{\beta ;\alpha } \right) 
	&= \frac{4}{3}\epsilon u_{\alpha }u^0  \\
 \frac{1}{2\sqrt{\gamma }}\partial _t 
 \left(\sqrt{\gamma } k_{\alpha }^{\beta }\right) 
+ P_{\alpha }^{\beta } 
  &= \frac{\epsilon}{3} \left( u_{\alpha }u^{\beta} 
+ {\delta }_{\alpha }^{\beta } \right) \, ,
\label{eeqil3}
\end{align}
\label{eeqil}
\eseq
where
$P_{\alpha }^{\beta } = 
{\gamma }^{\beta \gamma }P_{\alpha 
\gamma }$ represents the three-dimensional 
Ricci  tensor obtained by the 
metric ${\gamma }_{\alpha \beta}$
and $u_i $ $(i=0,1,2,3)$ denotes the matter 
four-velocity vector field. 

Let us consider a spatial metric of the form 
\beq
g_{\alpha \beta}= t~ a_{\alpha \beta}
+ t^2~ b_{\alpha \beta} + \ldots \, ,
\label{gqi}
\eeq
whose inverse reads as
\beq
g^{\alpha \beta}= t^{-1} a^{\alpha \beta}
- b^{\alpha \beta} + \ldots \, ,
\label{gqiinv}
\eeq
being $a^{\alpha \beta}$ the inverse tensor 
to $a_{\alpha \beta}$ which is the one used 
for the operations of rising and lowering 
indices as well as for the covariant 
differentiation.
Once (\ref{gqi}) is substituted in the field 
equations (\ref{eeqil}),  
we find the energy density and four-velocity 
to leading order as
\bseq
\bea
\epsilon &=& \frac{3}{4 t^2} - \frac{b}{2t} \\
u_{\alpha} &=& \frac{t^2}{2}
\left( b_{;\alpha} - b^{\beta}_{~\alpha ;\beta}\right) \, ,
\label{lkuu}\eea
\eseq
 respectively.
The three-dimensional Christoffel symbols and the 
tensor $P_{\alpha \beta}$ are, to first approximation, 
independent of time and (\ref{eeqil3}) gives
\beq
\label{eeqsol3}
P_{\alpha}^{~\beta} + \frac{3}{4} b_{\alpha}^{~\beta}+
\frac{5}{12}\delta_{\alpha}^{~\beta} b =0
\eeq
and then 
\beq
\label{eeqsol3b}
b_{\alpha}^{~\beta}= - \frac{4}{3} P_{\alpha}^{~\beta}+
\frac{5}{18}\delta_{\alpha}^{~\beta} P \, .
\eeq
The six functions $a_{\alpha \beta}$ are arbitrary
and, once they are given, the coefficients
$b_{\alpha \beta}$ are determined by (\ref{eeqsol3b})
and hence 
also the density of matter and its velocity can be 
derived. 
When $t\rightarrow 0$ the distribution of matter 
becomes homogeneous and its density approaches
a value which is coordinate independent.  
The expression giving the distribution of velocity
follows from (\ref{lkuu}) explicitly as
\beq
u_{\alpha} = \frac{t^2}{9}b_{;\alpha} \,.
\eeq

Such a  framework is completed considering 
that an arbitrary transformation of the spatial 
coordinates (for example to reduce $a_{\alpha \beta}$
to a diagonal form) leaves only three  
arbitrary functions allowed in this quasi-isotropic
solution, while the fully isotropic model is 
recovered in the specific choice of $a_{\alpha \beta}$
corresponding to the space of constant curvature
$P_{\alpha \beta}= \textrm{const.}
\times \delta_{\alpha \beta}$.

\section{Quasi-isotropic Solution Towards Singularity
with a Scalar Field
\label{quasimontani}}
%%%% da montani

In this Section we show how, following 
\cite{M99, M00},  
the quasi-isotropic Universe dynamics 
in presence of ultra-relativistic matter 
and a real self-interacting scalar field 
behaves in the asymptotic 
limit to the cosmological singularity,
while in the next Section, following
\cite{IM03} we will see the opposite 
limit far from the singularity, opening 
the way also to a 
generalization \cite{IM03a, IMmg10c}.  

In particular, the presence of the scalar 
field kinetic term allows the existence of a 
quasi-isotropic solution characterized by an 
arbitrary spatial dependence of the energy 
density associated with the 
ultra-relativistic matter. 
To leading order, there is no direct 
relation between the 
isotropy of the Universe and the homogeneity 
of the ultra-relativistic matter in it distributed.

%\end{abstract}

However, as shown in \cite{BKL70,BKL82} 
(see also  \cite{KK93}), 
the general behaviour of the Universe near the 
initial Big-Bang is characterized by a completely 
disordered dynamics and an increasing 
degree of anisotropy, up to develop a fully 
turbulent regime. 

Hence, the contrast between such a general 
tendency to anisotropy and the evidence 
that in the forward evolution since a given age
the Universe should have performed 
an highly symmetric behaviour is a  
problem related to properties 
of the Universe evolution 
at very different stages of anisotropy.

The quasi-isotropic 
solution  allows, far enough from 
the initial singularity, the oscillatory regime 
\cite{BKL70,BKL82} to be decomposed in terms of a
quasi-isotropic component plus suitable wave-like 
small corrections. 
An analogous decomposition has been obtained in 
\cite{GDY75} for the Bianchi type 
IX model as a homogeneous prototype of the general 
inhomogeneous case. 

Here we summarize the features acquired by a 
quasi-isotropic solution 
(\idest one in which the three spatial 
directions are dynamically equivalent) in presence of 
ultra-relativistic matter and a real self-interacting 
scalar field. 
Then a quasi-isotropic 
model solution exists and is characterized,
asymptotically to the Big Bang, by an 
arbitrary distribution of the 
ultra-relativistic matter and in which the 
spatial curvature component has no 
dynamical role to first two orders of 
approximation. 
The presence of the scalar field  kinetic term, 
close enough to the singularity, modifies
deeply the general cosmological solution
leading to the appearance of a dynamical 
regime characterized, point by point in space, by 
the monotonical collapse of the
three spatial directions \cite{BK73,KK87}. 

Let us consider a synchronous reference frame in 
which the 
line element reads as
\beq
ds^2 = c^2dt^2 - 
\gamma_{\alpha\beta}(t, x^\gamma)
dx^{\alpha }dx^{\beta } \, ,
\label{monta1}
\eeq
the matter is described by a perfect fluid with  
ultra-relativistic equation of state 
$p = \frac{\epsilon }{3}$ and the scalar field 
$\phi (t, x^{\gamma })$ admits a 
potential term $V(\phi )$; 
the Einstein equations reduce to the 
partial differential system
\bseq
\begin{align} 
&\frac{1}{2} \partial _t k_{\alpha }^{\alpha } 
		+ \frac{1}{4} 
		k_{\alpha }^{\beta }k_{\beta }^{\alpha } =  %\\
 		\chi \left[ - \left(4{u_0}^2 - 1 \right)\frac{\epsilon }{3} -  
		\frac{1}{2}(\partial _t\phi )^2 + V(\phi )\right] 
		\label{meeq1}\\
&\frac{1}{2}  \left(k^{\beta }_{\alpha ;\beta } 
		- k^{\beta }_{\beta ;\alpha }\right) = 
		\chi \left( \frac{4}{3}\epsilon u_{\alpha }u_0 + \frac{1}{c}
		\partial _{\alpha }\phi \partial _t \phi \right) 
		\label{meeq2}\\
&\frac{1}{2\sqrt{\gamma }} \partial _t \left(\sqrt{\gamma }
 		k_{\alpha }^{\beta }\right) + P_{\alpha }^{\beta } = 
 			\chi \left[ {\gamma }^{\beta \sigma }
		\left( \frac{4}{3}\epsilon u_{\alpha }u_{\sigma } 
		+ \partial _{\alpha }\phi \partial _{\sigma }\phi \right) +
		\left( \frac{\epsilon }{3} + 
		V(\phi )\right) {\delta }_{\alpha }^{\beta }\right] 
\label{meeq3}
\end{align}
\label{montaeeq}
\eseq
where, as usual, $\chi $ is the Einstein constant 
$\chi =\displaystyle \frac{8\pi G}{c^4}$ 
(we take $c=1$),
obvious notation for derivatives. 

The partial differential equation describing the 
scalar field $\phi (t, x^{\gamma })$
dynamics, deeply coupled to the 
Einstein ones, reads as 
\beq
\partial_{tt}\phi + \frac{1}{2}k_{\alpha }^{\alpha }
\partial_t\phi - 
{\gamma }^{\alpha \beta }{\phi }_{;\alpha \beta } 
+ \frac{dV}{d\phi } = 0 
\label{montaefi}
\eeq
and finally the hydrodynamic equations
accounting for the matter evolution
are explicitly \cite{LK63} 
\bseq
\begin{align}
& 	\frac{1}{\sqrt{\gamma }}
		\partial _t\Big(\sqrt{\gamma }
		{\epsilon 	}^{3/4}u_0\Big) 
 			+ \frac{1}{\sqrt{\gamma }}
		\partial_{\alpha }(\sqrt{\gamma }{\epsilon }^{3/4}
		u^{\alpha }) = 0  
		\label{or}\\
& 4 {\epsilon }\left( \frac{1}{2}\partial _t{u_0}^2 
		+ u^{\alpha }
		\partial _{\alpha }u_0 + 
		\frac{1}{2}k_{\alpha \beta }u^{\alpha }
		u^{\beta }\right) = %\nonumber \\
%& \qquad \qquad\qquad \qquad 	= 
\left(1 - {u_0}^2\right)
		\partial _t\epsilon - u_0u^{\alpha }
		\partial _{\alpha }\epsilon 
		\label{pr}\\
%\begin{align}
& 4\epsilon \left( u_0\partial _tu_{\alpha } 
		+ u^{\beta }
		\partial _{\beta }u_{\alpha } 
		+ \frac{1}{2}u^{\beta }u^{\gamma }
		\partial_{\alpha }{\gamma }_{\beta \gamma }\right) = 
			%	\nonumber \\
		%&\qquad \qquad \qquad \qquad	= 
- u_{\alpha }u_0\partial _t\epsilon +
 		\Big({\delta }_{\alpha }^{\beta } 
 		- u_{\alpha }u^{\beta }\Big)
\partial_{\beta }\epsilon  \,.
\label{qr}
\end{align}
\label{montahydro}
\eseq

Any kind of matter described by a perfect 
fluid energy-momentum tensor with equation 
of state $p = w\epsilon $ ($w$ is a constant), 
$w \neq 0$,
is dynamically equivalent to a scalar field 
$\psi (t, x^{\gamma })$ 
with Lagrangian density  
\beq
{\cal L} = \frac{1}{2}\sqrt{-g} 
\left( g^{ik}\partial _i\psi \partial _k\psi \right) 
^{\frac{1}{2}\left( \frac{1}{{\it w}} + 1 \right) }
\eeq 
once identified 
\bseq
\bea
\epsilon &\equiv& \frac{1}{2w}
\left( g^{ik}\partial_i\psi \partial_k\psi 
\right)^{\frac{1}{2}\left( \frac{1}{w} + 1 \right) }
\, , \\ 
p &\equiv& \frac{1}{2}\left( g^{ik}\partial _i\psi \partial _k\psi \right)^{\frac{1}{2}\left( \frac{1}{w} + 1 \right) } \, ,\\
u_i &\equiv& \frac{\partial _i \psi }{ 
\sqrt{g^{ik}\partial _i\psi \partial _k\psi }} \, ,
\eea
\eseq
where $g_{ik}$ $(i,k=0,1,2,3)$ is the 
four-dimensional covariant metric.
The considered (Klein-Fock) scalar field $\phi $ 
($w = 1$) corresponds to a 
perfect fluid with equation of state $p = \epsilon $, 
as well as the 
ultra-relativistic matter 
($p = \frac{\epsilon }{3}$) is dynamically 
equivalent to a scalar field 
$\psi $, described by the above Lagrangian density
in the case $w = \frac{1}{3} $. \\
The Einstein equations follow by
the  variational principle
\beq 
\delta S = \delta \int \sqrt{- g}\left\{ R - \chi \left[  
g^{ik}\partial _i\phi \partial _k\phi + \left( g^{ik}\partial _i\psi 
\partial _k\psi \right) ^2\right] \right\} d^4x =0
\eeq
where $R$ is the four-dimensional curvature scalar.  \\
The quasi-isotropic solution 
near the cosmological 
singularity,  
as seen in Section \ref{quasiisoKL} 
(and in \cite{LK63}), 
  refers to a Taylor expansion
of the three-dimensional metric 
time dependence as (see \ref{gqi}) 
\beq 
\gamma_{\alpha \beta}(t, x^{\gamma }) = \sum_{n=0}^{\infty }
{a^{(n)}}_{\alpha \beta } 
(x^{\gamma })\left( \frac{t}{t_0}\right) ^n 
\eeq 
where
\beq
a^{(n)}_{\alpha \beta }(x^{\gamma }) \equiv 
\left. \frac{{\partial }^n{\gamma }_{\alpha \beta }}{{\partial }t^n}\right|_{t=0}{t_0}^n  
\eeq 
in which $t_0 $ is an arbitrarily fixed instant of time 
($t\ll t_0$) and the existence of the singularity
implies $a^{(0)}_{\alpha \beta }\equiv 0$. 
In \cite{LK63} only 
the first two terms appear, \idest 
\beq
{\gamma }_{\alpha \beta } = 
a^{(1)}_{\alpha \beta }\frac{t}{t_0} + 
a^{(2)}_{\alpha \beta }\left(\frac{t}{t_0}\right)^2 \, .
\eeq
The presence of the scalar field permits to 
relax the assumption of  
expandability in integer powers.\\ 
In order to introduce in a quasi-isotropic scenario
(eventually inflationary, see below Section
\ref{quasiiinflas}) 
 small inhomogeneous 
corrections to leading order, we require a 
three-dimensional metric tensor having the 
following structure 
\begin{align}
{\gamma }_{\alpha \beta }\left(t, x^{\gamma }\right) &= 
a^2(t){\xi }_{\alpha \beta }\left(x^{\gamma }\right) + 
b^2(t){\theta }_{\alpha \beta }\left(x^{\gamma }\right) 
+ \mathcal{O}\left(b^2\right) = \nonumber \\
&=  a^2(t)\Big[ {\xi }_{\alpha \beta }\left(x^{\gamma }\right) + 
\eta (t){\theta }_{\alpha \beta }\left(x^{\gamma }\right) 
+ \mathcal{O}\left(\eta^2 \right)\Big] \, ,
\label{sr} 
\end{align}
where we defined 
$\displaystyle\eta \equiv \frac{b^2}{a^2}$  
and suppose $\eta$ to satisfy the 
condition 
\beq 
\lim_{t\rightarrow \infty} \eta (t) = 0 \,.
\label{tr}
\eeq 
In the limit of the considered 
approximation, the inverse three-metric reads 
\beq
{\gamma }^{\alpha \beta }\left(t, x^{\gamma }\right) = 
\frac{1}{a^2(t)}\Big( 
{\xi }^{\alpha \beta }\left(x^{\gamma }\right) -  
\eta (t){\theta }^{\alpha \beta }\left(x^{\gamma }\right) 
+ \mathcal{O}\left(\eta^2 \right)\Big)  \, ,
\label{ss}
\eeq
where ${\xi }^{\alpha \beta }$ denotes the 
inverse matrix of ${\xi }_{\alpha \beta }$ 
and assumes a metric role, \idest we have 
\beq 
{\xi }^{\beta \gamma }{\xi }_{\alpha \gamma } = 
{\delta }_{\alpha }^{\beta } \, ,\qquad 
{\theta }^{\alpha \beta } = 
{\xi }^{\alpha \gamma }{\xi }^{\beta \delta }
{\theta }_{\gamma \delta } \, .
\label{ortt} 
\eeq
The covariant and contravariant 
three-metric expressions lead to the 
important explicit relations
\beq
k_{\alpha }^{\beta } = 
2\frac{\dot{a}}{a}{\delta }_{\alpha }^{\beta } + 
\dot{\eta }{\theta }_{\alpha }^{\beta } \quad \Rightarrow \quad 
k_{\alpha }^{\alpha } = 6\frac{\dot{a}}{a} + \dot{\eta }\theta 
\,, \qquad \theta \equiv {\theta }_{\alpha }^{\alpha } \, .
\label{ur} 
\eeq
Since the fundamental equality 
$\partial_t(\ln \gamma ) = k_{\alpha }^{\alpha }$
holds, 
then we immediately get
\begin{align} 
\gamma = j
			a^6e^{\eta \theta } \quad \Rightarrow \quad 
	\sqrt{\gamma } &= 
				\sqrt{j}a^3e^{\frac{1}{2}\eta \theta } 
												\sim \nonumber \\
&\sim \sqrt{j}
		a^3\left( 1 + \frac{1}{2}\eta \theta + 
		\mathcal{O}(\eta^2 )\right) \,, 
\label{v} 
\end{align}
once defined $j\equiv \det{\xi }_{\alpha \beta }$.

The Landau-Raychaudhury theorem \cite{LF} 
applied to the 
present case implies the condition
\beq 
\lim_{t\rightarrow 0}a(t) = 0   \, .
\eeq

The set of field equations (\ref{montaeeq}) 
is solved retaining only 
the terms linear in $\eta $ and its time 
derivatives and neglecting all terms containing 
spatial derivatives of the dynamical variables, 
in order to obtain 
asymptotic solutions in the 
limit $t\rightarrow 0$ and finally  
checking the self-consistence of the 
approximation scheme. 
The possibility to neglect the potential term 
$V(\phi )$ is not ensured by the 
field equations but is based on the idea 
that, in an inflationary scenario, the 
scalar field potential energy becomes  
dynamically relevant only during the 
``slow-rolling phase'', far from the 
singularity, while  the 
kinetic term asymptotically dominates. 
With this in mind the solution for $a(t)$ is found 
to be
\beq
a(t) = \left( \frac{t}{t_0}\right)^{\frac{1}{3}} \, ,
\label{montaat} 
\eeq 
where $t_0$ is an integration constant, while
for $\eta(t)$ it is
\beq 
\eta (t) = \left( \frac{t}{t_0}\right) ^{\frac{2}{3}} \, ,
\label{montaeta} 
\eeq 
and for $u_{\alpha}$
\beq
u_{\alpha }(t, x^{\gamma }) = v_{\alpha }(x^{\gamma })
\left( \frac{t}{t_0}\right) ^\frac{1}{3} + 
\mathcal{O}\left( \frac{t}{t_0}
\right)  \, ,
\eeq
respectively; hence we get 
for the tensor 
${\theta }_{\alpha \beta }(x^{\gamma })$
the expression 
\beq 
{\theta }_{\alpha \beta } = \frac{\rho }{3 + 4v^2}
\Big[ 
\left(1 - 2v^2\right){\xi }_{\alpha \beta } 
+ 10v_{\alpha }v_{\beta }\Big]  
\quad \Rightarrow \quad \theta = \rho \, ,
\label{montath} 
\eeq
where $\rho(x^{\gamma })$ denotes an 
arbitrary function of the spatial 
coordinates. \\
The energy density of the ultra-relativistic 
matter, to leading order, 
has the  expression
\beq 
\epsilon (t, x^{\gamma }) = \frac{5\rho (x^{\gamma })}
{ 3\chi\Big[3 + 4v^2(x^{\gamma })\Big]{t_0}^{2/3}t^{4/3}} 
+ \mathcal{O}\left( \frac{t}{t_0} \right) \, , 
\label{montaeps} 
\eeq 
and, once integrated the scalar field 
equation (\ref{montaefi}), they provide 
\begin{align} 
&\phi (t, x^{\gamma }) = 
\sqrt{\frac{2}{3\chi }}
\left[ \ln\left( \frac{t}{t_0}\right) - 
\frac{3}{4} 
\left( \frac{t}{t_0}\right) ^{\frac{2}{3}}
\rho (x^{\gamma }) + 
\sigma (x^{\gamma }) \right] 
+ \mathcal{O}\left( \frac{t}{t_0} \right) 
\label{montafisol}
\end{align}
where $\sigma (x^{\gamma })$ 
is an arbitrary function of the spatial 
coordinates. \\
Finally, equation (\ref{meeq3})
yields the expression for the 
functions $v_{\alpha }$ in terms of 
$\rho $ and of the spatial gradient 
$\partial _{\alpha }\sigma $ as
\bseq
\bea 
v_{\alpha } &=& - 
\frac{3(3 + 4v^2)}{10\rho \sqrt{1 + v^2}}t_0
\partial_{\alpha } \sigma \, , \\ 
v^2 &=& \frac{24{\tau }^2 - 1 + 
\sqrt{1 - 12{\tau }^2}}{2(1 - 16{\tau }^2)} 
\eea
where $\tau$ represents the quantity
\beq 
\tau \equiv \frac{3t_0}{10\rho }
\sqrt{{\xi }^{\alpha \beta }
\partial_{\alpha }\sigma \partial_{\beta }\sigma } \, .
\eeq 
\eseq
The particular  and simple case 
$\sigma \equiv 0$, in correspondence to 
which $v^2 \equiv 0$ $(v_{\alpha } \equiv 0)$ 
leads to the solutions 
\bseq
\begin{align}
&{\theta }_{\alpha \beta } = 
			\frac{1}{3}\rho (x^{\gamma })
			{\xi }_{\alpha \beta } \\
&\epsilon (t, x^{\gamma }) = 
		\frac{5}{9\chi }\rho (x^{\gamma })
		\frac{1}{{t_0}^{2/3}t^{4/3}} 
		+ \mathcal{O} \left( \frac{t}{t_0} \right)  \\
&\phi (t, x^{\gamma }) = 
		\sqrt{\frac{2}{3\chi }}\ln
		\left( \frac{t}{t_0}\right) - 
		\frac{3}{4} 
		\sqrt{\frac{2}{3\chi }} 
		\left( \frac{t}{t_0}\right)^{2/3}
		\rho (x^{\gamma })  
		+ \mathcal{O}\left( \frac{t}{t_0} \right) \\
& u_{\alpha }(t, x^{\gamma }) = 
		\frac{3}{8}\partial _{\alpha }
			\ln\big( \rho (x^{\gamma })\big) t 
		+ \mathcal{O}\left( \frac{t}{t_0} \right) \, .
\end{align} 
Finally we obtain the three-dimensional 
metric tensor as 
\begin{align}
&\!\! {\gamma }_{\alpha \beta }(t, x^{\gamma }) = 
\left( \frac{t}{t_0}\right) ^{2/3}\left[ 
1 + 
\left( \frac{t}{t_0}\right) ^{2/3}
\frac{\rho (x^{\gamma })}{3}\right]
{\xi }_{\alpha \beta }(x^{\gamma }) 
+ \mathcal{O}\left( \frac{t}{t_0} \right) \,.
\end{align}
\label{montatuttes}
\eseq
On the basis of equations  (\ref{montatuttes}),
the hydrodynamic ones (\ref{montahydro}) 
reduce to an identity 
in the considered approximation. 

The solution here shown is completely 
self-consistent up to first-two orders 
in time and contains five physically 
arbitrary functions of the spatial coordinates:
three out of the six functions 
${\xi }_{\alpha \beta }$ 
(the remaining three of them can be fixed by 
pure spatial 
coordinates transformations), 
the spatial scalar $\rho (x^{\gamma })$ 
and the function $\sigma (x^{\gamma })$.  

The independence among the functions 
${\xi }_{\alpha \beta }$, $\rho $ and $\sigma $ 
implies the existence of a quasi-isotropic dynamics 
in correspondence to an arbitrary spatial 
distribution of ultra-relativistic matter. \\
The kinetic term of the scalar field 
behaves, to leading order, 
as $\sim \displaystyle a^{-6}$ and therefore, 
in the limit $a\rightarrow 0$, 
dominates over the ultra-relativistic energy 
density; in fact, the latter term diverges only as 
$\sim \displaystyle a^{-4}$  and
the spatial curvature as  
$\sim \displaystyle a^{-2}$, making them 
negligible. 

Furthermore, for a generic equation of state 
$p = w\epsilon $ the corresponding matter 
energy density behaves asymptotically as  
$\sim a^{-3(1 + w)}$,
but the above dynamical scheme is not 
qualitatively affected when considering 
values of $w$ in the range 
$-\displaystyle \frac{1}{3} < w < 1$
instead of the ultra-relativistic case 
$w =\displaystyle \frac{1}{3}$.

\section{Quasi-isotropic Inflationary Solution} 
%
%
%%%  da quasiisotropic inflation ijmpd03

In this Section we find a solution 
for a quasi-isotropic inflationary 
Universe which allows to introduce in the 
problem a certain degree of 
inhomogeneity \cite{IM03}. 
We consider a model which generalizes 
the (flat) FLRW one by introducing a 
first-order inhomogeneous term, whose 
dynamics is induced by an effective 
cosmological constant. 
The three-metric tensor consists  
of a dominant term, corresponding to 
an isotropic-like component, while 
the amplitude of the first-order one 
is controlled by a ``small''  
function $\eta(t)$. 

In a Universe filled with ultra-relativistic
matter and a real self-interacting scalar 
field, we discuss the resulting dynamics, 
up to first order in $\eta$, when the scalar 
field performs a slow roll on 
a plateau of a symmetry breaking 
configuration and induces an effective 
cosmological constant.

We show how the spatial distribution 
of the ultra relativistic matter and of 
the scalar field admits an arbitrary 
form but nevertheless, due to the required
inflationary e-folding, 
it cannot play a serious dynamical
role in tracing the process of structures 
formation (via the Harrison--Zeldovich 
spectrum). 
As a consequence, we find 
reinforced the idea that the inflationary 
scenario is incompatible with a classical
origin of the large scale structures.

\subsection{Quasi-isotropic Inflation 
and Density Perturbation}

As we have seen, the inflationary model 
\cite{G81, CW73}  
is, up to now,
the most natural and complete scenario to make
account of the problems outstanding in the
Standard Cosmological Model, 
like the horizons and flatness 
paradoxes \cite{KT90} 
(for pioneer works on inflationary 
scenario and the spectrum of gravitational 
perturbation, see also \cite{STA80,STA79}); 
indeed such
a dynamical scheme on one hand is able to justify
the high isotropy of the cosmic microwaves 
background radiation
(characterized by temperature fluctuations
$\mathcal{O}(10^{-4})$ \cite{dB01}) and, on the other one, 
provides a mechanism for generating 
a (scale invariant) 
spectrum of inhomogeneous perturbations
(via the scalar field quantum fluctuations).\\ 
Moreover, as shown in \cite{KM02,Star83}, a
slow-rolling phase of the scalar field allows 
to connect the generic inhomogeneous Mixmaster 
dynamics \cite{BKL70, IM02}
with a later quasi-isotropic Universe
evolution, in principle 
compatible with the actual cosmological
 picture, \cite{T95}.\\
With respect to this,  we investigate the 
dynamics performed by small inhomogeneous
 corrections to a leading order metric, 
during inflationary expansion.

The model presented in \cite{IM03} has the relevant
feature to contain inhomogeneous corrections
to a flat FLRW Universe, which in principle 
could take a role to understand the process of 
structure formation, even in presence of an 
inflationary behaviour;
however, a careful analysis of our result prevents
this possibility in view of the strong inflationary
e-folding, so confirming the expected incompatibility 
between an inflationary scenario and a classical origin
of the Universe clumpyness. 

In what follows, we will use the 
quasi isotropic solution
which was introduced in \cite{LK63} 
(see Section  \ref{quasiisoKL})
 as the 
simplest, but rather general, extension of the 
FLRW model; for a discussion of the quasi isotropic 
solution in the framework of the 
``long-wavelength'' approximation see \cite{DT94},
while for the implementation of such a solution
after inflation \cite{K8385} 
to a generic equation of state
and to the case of two ideal hydrodynamic fluid
see \cite{KHA02} and \cite{KHA03}, respectively. \\
In the previous Section \ref{quasimontani} 
this solution was discussed in the presence
of a real scalar field kinetic energy, 
leading to a power-law solution for the three-metric, 
and predicted interesting features for the 
ultra-relativistic matter dynamics. 

We analyse here
the opposite dynamical scheme, \idest when the scalar 
field undergoes a slow-rolling phase since the 
effective cosmological constant dominates its 
kinetic energy.
We provide, 
up to first-two orders of approximation 
and in a synchronous reference,
a detailed description of
the three-metric, of the scalar field
and of the ultra relativistic matter dynamics,
showing that the
volume of the Universe expands exponentially 
and induces a corresponding exponential decay 
(as the inverse fourth power of the cosmic 
scale factor), either of the three-metric 
corrections, as well as of the 
ultra relativistic matter
(the same behaviour characterizes roughly even 
the scalar field inhomogeneities).
It is remarkable that the spatial dependence 
of such component is described by a function 
which remains an arbitrary degree of freedom; 
in spite of such freedom in fixing the primordial 
spectrum of inhomogeneities, 
due to the inflationary e-folding,
we show there is no chance that,
after the de Sitter phase,
such relic perturbations can survive enough 
to trace the large scale
structures formation by an 
Harrison--Zeldovich spectrum.\\
This behaviour suggests that the spectrum of
inhomogeneous perturbations \cite{MB95} cannot 
arise  directly by the 
classical field nature, but by its 
quantum dynamics.

Finally, we recall that the presence of 
the kinetic term of a scalar field,
here regarded as negligible, 
induces, near enough to the singularity, 
a deep modification of the general cosmological 
solution, leading to the appearance of a dynamical 
regime, during which 
the three spatial directions 
behave monotonically \cite{BK73,B99}
point by point in space.

%%%%%%%%%%%%%%%%%%%%%%%%%%%%%%%%%%%%%inizio parte inserita luglio2003

\subsection{Inhomogeneous Perturbations 
from an Inflationary Scenario}

The theory of inflation is based on the idea 
that during the Universe evolution a phase 
transition takes place (for instance 
associated with a spontaneous symmetry breaking 
of a Grand Unification model of strong and 
electroweak interactions) which induces an 
effective cosmological constant dominating 
the expansion dynamics. As a result, an exponential 
expansion of the Universe arises  and, under a 
suitable fine-tuning of the parameters, 
it is able to ``stretch'' so strongly the 
geometry that the \textit{horizon} and
\textit{flatness paradoxes} of the  SCM 
are naturally solved.\\
In the \textit{new inflation} theory 
(see Section \ref{slowrolling}), the Universe
undergoes a de Sitter phase
when the scalar field performs a ``slow-rolling'' 
behaviour over a very flat region of the potential
between the false and true vacuum.
The exponential expansion ends with the
scalar field falling down in the potential well
associated to the real vacuum and the scalar 
field dies via damped (by the expansion of the 
Universe and particles creation) 
oscillations which reheat the cold Universe left by the
de Sitter expansion (the relativistic particles
temperature is proportional to the inverse scale factor). 
Indeed, the decay of this super-cooled bosons condensate
into relativistic particles 
-- as a typical irreversible process -- 
generates a huge amount of entropy, which allows 
to account for the present high value 
$(\sim \mathcal{O}(10^{88}))$
of the Universe entropy per
comoving volume.

Apart from the transition across the
potential barrier between false and true vacuum,
which takes place in general via a tunnelling, 
the whole inflationary dynamics can be 
satisfactorily described via a classical 
\textit{uniform} scalar field $\phi = \phi (t)$.
The assumption that the field behaves in a classical 
way is supported by its bosonic and cosmological 
nature, but the existence of quantum fluctuations 
of the field within the different inflationary 
``bubbles'' leads to relax the hypothesis  of
dealing with a perfectly uniform scalar field.

In general, when analysing density perturbations,
it is  convenient to introduce the dimensionless
quantity \cite{KT90, Pad}
\begin{equation}
\delta\rho(t, x^{\gamma })\equiv 
\frac{\Delta\rho(t, x^{\gamma })}{\bar\rho}=
\frac{\rho - \bar\rho}{\bar\rho}
\, , 
\label{xax}
\end{equation}
where $\bar\rho$ denotes the mean density 
and $\gamma=1,2,3$.
The best formulation of the density perturbations theory           
is obtained expanding $\delta \rho$ 
in its Fourier components, or modes, 
\begin{equation}
\delta\rho_k=\frac{1}{(2\pi )^{3}}   \int d^3x \,e^{ik_{\alpha }x^{\alpha }}
\delta\rho(t, x^{\gamma })
\, .
\label{xax1}
\end{equation}
As long as the perturbations are in
the linear regime, \idest $\delta\rho_k\ll 1$,
it is possible to follow appropriately the dynamics 
of each mode with wave number 
$k$, which corresponds to a wavelength
$\displaystyle\lambda=\frac{2\pi}{k}$;
however, the
\emph{physical} size
of the perturbations  in an expanding Universe 
evolves via the 
\emph{cosmic scale factor} which from now on
we write as $a(t)$, in order to 
avoid ambiguities.

Since in the Standard Cosmological Model,
the ``Hubble radius''
scales as $H^{-1}\propto t$, while $a(t)\propto t^n$
with $n<1$, then every
perturbation, now inside the Hubble radius, 
was outside it at some earlier time.
We stress how 
the perturbations with a physical size
smaller or greater than the Hubble radius
have a very different dynamics, respectively, 
the former ones being affected by the action 
of the microphysics processes.

In the case of an inflationary scenario,
the situation is quite different. 
Since during the de Sitter phase 
the Hubble radius remains constant
while the cosmic scale
factor ``blows up'' exponentially, 
hence all cosmologically interesting 
scales have crossed the horizon
twice, \idest  
the perturbations begin sub-horizon sized,
cross the Hubble radius
during inflation and later cross back again inside
the horizon. \\
This feature has a strong implication on the
initial spectrum of density perturbations
predicted by inflation.
We present a qualitative argument to understand
how this spectrum can be generated. 

During inflation, the density perturbations
are expected to arise from the quantum mechanical
fluctuations of the scalar field $\phi$; 
these are, as usual, decomposed in their Fourier components
$\delta\phi_k$, \idest
\begin{equation}
\delta\phi_k=\frac{1}{(2\pi )^{3}} 
\int d^3x\,e^{ik_{\alpha }x^{\alpha }}
\delta\phi (t, x^{\gamma })
\, .
\label{xax2}
\end{equation}
The spectrum of quantum mechanical fluctuations of the
scalar field is defined as
\begin{equation}
(\Delta\phi)^2_k\equiv
\frac1{\cal V}\frac{k^3}{2\pi^2}\left|\delta\phi_k\right|^2
\label{xax3}
\, ,
\end{equation}
where ${\cal V}$ denotes the comoving volume.
For a massless minimally coupled scalar field in a de Sitter
space-time,
which approximates very well the real physical situation during
the Universe exponential expansion, it is well known that
(see \cite{KT90}) 
%
%\S 8.4 and references therein)
%
\begin{equation}
(\Delta\phi)^2_k=\left(\frac{H}{2\pi}\right)^2
\label{xax4}
\, ,
\end{equation}
then the mean square fluctuation of $\phi$
 takes the form
 \begin{equation}
	(\Delta\phi)^2=\frac1{(2\pi)^3V}\int d^3k
	\left|\delta\phi_k\right|^2=
	\int\left(\frac{H}{2\pi}\right)^2d(\ln k)
\label{xax5}
\, .
\end{equation}
Since $H$ is a constant during the de Sitter phase, 
each mode $k$ contributes roughly
the same amplitude to the mean square fluctuation. 
Indeed, 
the only dependence on $k$ takes place in 
the logarithmic term, but  
the modes of cosmological interest lay 
between $1$ Mpc and $3000$ Mpc
(it is commonly adopted the convention to set
the actual cosmic scale factor equal to unity), 
corresponding to a logarithmic interval
of less than an order of magnitude.\\
Thus we can conclude that
any mode $k$ crosses the horizon
having almost a constant amplitude
$\delta\phi_k\simeq H/2\pi$.
A delicate question concerns the mechanism by which
such quantum fluctuations of the scalar field achieve
a classical nature
%(for a detailed 
%discussion see 
\cite{STA96}; 
here we simply observe how
each mode $k$, once reached a classical stage,
is governed by the dynamics
\begin{equation}
 \delta\ddot \phi_k+3H\delta\dot\phi_k +
 \frac{ k^2}{a^2} {\delta\phi_k}=0 \, ; 
\label{xax6}
\end{equation}
%\fbox{label{mot}-tolta}
according to this equation, 
 super-horizon modes $k\ll a H$
(\idest $\lambda _{phys}\gg H^{-1}$) 
admit the trivial dynamics (\ref{xax6}) with
$\delta\phi_k\sim \mathrm{const.}$.
This simple analysis implies the important 
feature that any mode re-enters the horizon 
with roughly the same amplitude it had at 
the first horizon crossing. 
The spectrum of perturbations so generated 
is then induced into the relativistic energy 
density coming from the reheating phase, 
associated with the bosons decay; since then on,  
the evolution of the perturbation spectrum 
follows a standard paradigm.

The density perturbations discussed so far 
are related to the scalar field by
	\begin{equation}
	\delta\rho=\frac{\partial V}{\partial \phi}\delta\phi \, ,
	\end{equation}
where, because of the potential is very flat 
during inflation,  $\partial V/ \partial\phi$
is approximatively constant and then 
	\begin{equation}
	\delta\rho\simeq\mathrm{const.}\times \delta\phi \, .
	\end{equation}
The spectrum of density perturbations has a 
Harrison-Zeldovich form, characterized by a
 constant amplitude: this is a very generic 
 prediction of inflation,  based
on the features of the potential flatness
common to nearly all inflationary models. 
On the other
hand, the spectrum amplitude is model dependent
and accurate measures could discriminate
between the various models.

Finally, we discuss the Gaussian distribution
of the quantum mechanical fluctuations
generated by the inflation: as long as
the field $\phi$ is minimally coupled, it has a
low self interaction and
each mode fluctuates independently; hence, 
since the fluctuations we actually observe are
the sum of many of the quantum
ones, their distribution can be expected 
to be Gaussian (as it should
be for the sum of many independent variables).\\
This is reflected  on the distribution of 
temperature fluctuations of the CMBR
as a powerful  test inflation. 
In a detailed analysis by \cite{Wu},
%Wu et al. 
82 (even if not independent) 
hypothesis tests for Gaussianity are implemented, 
showing how the MAXIMA map is consistent with Gaussianity 
on angular scales between 10' and $5^\circ$,
where deviations are most likely to occur. 

%%%%%%%%%%%%%%%%%%%%%%%%%%%%%%%%%%%%%fine parte inserita luglio2003

\subsection{Geometry, Matter and Scalar Field Equations}

When matter is represented as a perfect fluid with  
ultra relativistic equation of state 
$\displaystyle p = \frac{\epsilon }{3}$ 
 and in presence of a scalar field $\phi (t, x^{\gamma })$ 
with a potential term $V(\phi )$ the Einstein 
equations read as
\beq
\label{eet}
R_i^k = \chi \sum_{(z)=m,\phi}{\left({T_i^k}^{(z)} -
 \frac{1}{2}\delta_i^k {T_l^l}^{(z)} \right)}  
\eeq
where 
${T_i^k}^{(m)}$ and ${T_i^k}^{(\phi)}$ indicate, 
respectively, the energy-momentum tensor of
the matter and the scalar field, \idest
explicitly as in (\ref{montaeeq});
%	\label{br}  \\
%	\label{cr} \\
%	\label{dr} 
this set is coupled to 
the dynamics of the scalar field 
$\phi (t, x^{\gamma })$ given by (\ref{montaefi})
%\label{mr} 
and to 
the hydrodynamic equations (\ref{montahydro}),
which take into account for the matter 
evolution.
In view of the chosen feature for (\ref{eet}), 
equation (\ref{or}) doesn't contain any spatial 
gradients of the three-metric tensor and of the 
scalar field.
This scheme is completed by observing how it can
be made covariant with respect to coordinate 
transformations of the form 
\beq 
t^{\prime } = t + f(x^{\gamma }) \,, \qquad  
x^{\alpha \prime } = x^{\alpha \prime }(x^{\gamma }) 
%\label{r} 
\eeq
$f$ being a generic space dependent function.

\section{Quasi Homogeneous Inflationary Solution
\label{quasiiinflas}}

In order to introduce in a quasi isotropic (inflationary) 
scenario small inhomogeneous 
corrections to the leading order, we require a 
three-dimensional metric tensor having the 
structure as in (\ref{sr})-(\ref{ortt})
and (\ref{ur})-(\ref{v}).

The field equations (\ref{montaeeq})
%(\ref{meeq1})-(\ref{meeq3}) 
are analysed 
via the standard procedure of 
constructing asymptotic solutions in the limit 
$t\rightarrow \infty$, retaining only 
terms linear in $\eta$ and its time 
derivatives and by verifying {\it a posteriori} 
the self-consistency of the approximation scheme, 
\idest the neglected 
terms to be really of higher order in time 
\cite{IM03}.

%\subsection{Inflationary Solution}

In the quasi-isotropic approach,
we assume that the scalar field dynamics 
in the plateau region is governed by a 
potential 
\beq
V(\phi) = \Lambda + K(\phi) \, , 
\qquad \Lambda={\rm const.}
\label{w}
\eeq
where $\Lambda$ is the dominant term
and $K(\phi)$ is a small correction to 
it. The role of $K$, as shown in the 
following, is to contain
inhomogeneous 
corrections via the $\phi$-dependence;
the functional form of $K$ can
be any one of the most common inflationary 
potentials, 
as they appear near the flat 
region for the evolution of $\phi$.\\
What follows remains valid, for example, 
for the 
relevant cases of the quartic and 
Coleman--Weinberg 
expressions already introduced in some 
detail in Section \ref{slowrolling}
\beq
\label{potcol}
K(\phi)=\left\{
                \begin{array}{l} \displaystyle -\frac{\lambda}{4} \phi^4 \, ,  \qquad\qquad\qquad \qquad \lambda={\rm const.} \\
                \displaystyle B\phi^4\left[\ln\left(\frac{\phi^2}{\sigma^2}\right)-\frac{1}{2} \right] \, , \qquad \sigma={\rm const.} \, ,
                \end{array}
                \right. 
\eeq
viewed as corrections to the constant $\Lambda$ 
term, although explicit calculations are 
developed below only for the first case. 

Our inflationary solution is obtained under the 
standard requirements
\bseq
\begin{align}
\label{infla1}
&\frac{1}{2}\left(\partial_t \phi\right)^2 \ll 
	V\left(\phi\right) \\
& \mid \partial_{tt}\phi \mid \ll 
	\mid k^{\alpha}_{\alpha}\partial_t\phi \mid
\label{infla2} \, .
\end{align}
\label{infla12}
\eseq
The approximations (\ref{infla12}) 
and the substitution of 
(\ref{ur}) reduce the scalar field 
equation (\ref{montaefi}) to  
\beq
\left( 3\frac{\dot{a}}{a} + 
	\frac{1}{2}\dot{\eta }\theta \right) 
	\partial _t\phi - \lambda \phi^3= 0 
\label{xba} 
\eeq
where the contribution 
of the $\phi$ spatial gradient is assumed 
to be negligible.

Similarly, the quasi-isotropic approach 
(in which the inhomogeneities become relevant
only for the next-to-leading order), 
neglecting the spatial derivatives
in (\ref{or}), leads to 
\begin{align} 
\sqrt{\gamma }{\epsilon }^{3/4}u_0 = 	l(x^{\gamma })
 			%& 
 			\quad		\Rightarrow \quad %\nonumber \\
			%& \; \Rightarrow 
							\epsilon \sim 
\frac{l^{4/3}}{j^{2/3}a^4{u_0}^{4/3}} 
\left( 1 - \frac{2}{3}\eta \theta + 
\mathcal{O}(\eta^2 )\right) \ , ,
\label{a1} 
\end{align} 
where $l(x^{\gamma })$ denotes an arbitrary 
function of the 
spatial coordinates. 

Let us now face
in the same approximation scheme 
the analysis of the Einstein
equations (\ref{montaeeq}). 
Taking into account (\ref{infla1}),
equation (\ref{meeq1}) to first order in $\eta$ 
reads  
\beq
3\frac{\ddot{a}}{a} + 
\left[ \frac{1}{2}\ddot{\eta } + 
\frac{\dot{a}}{a}\dot{\eta } \right]
\theta - \chi \Lambda= - \chi 
\frac{\epsilon }{3}\left(3 + 4u^2 \right) 
\label{a2} 
\eeq
having set 
\beq 
u^2 \equiv \frac{1}{a^2}{\xi }^{\alpha \beta }
u_{\alpha }u_{\beta } 
\quad \Rightarrow \quad u_0 = \sqrt{1 + u^2}  \, .
\label{a3in} 
\eeq
Equation (\ref{meeq3}) reduces to the form 
\begin{align}
 \frac{2}{3}\big(a^3\big)_{,tt}~
	{\delta }_{\alpha }^{\beta } + 
&	\big(a^3 ~\eta_{,t} \big)_{,t}~
	{\theta }_{\alpha }^{\beta } + 
	\frac{1}{3}\Big[\big(a^3\big)_{,t}~\eta \Big]_{,t}~
	\theta {\delta }_{\alpha }^{\beta } +
		aA_{\alpha}^{\beta}= \nonumber \\
&= \chi \Big[ \frac{1}{a^2} \Big( {\xi }^{\beta \gamma } -
		\eta {\theta}^{\beta \gamma } \Big)
		\frac{4}{3}\epsilon u_{\alpha }u_{\gamma }+
		%\nonumber \\
%& \qquad \qquad \qquad + 
\left(\frac{\epsilon}{3} 
		+\Lambda \right){\delta }_{\alpha }^{\beta }\Big] 
		2a^3\left( 1 + \frac{\eta \theta}{2} \right)  \, ,
\label{a4} 
\end{align}
where we adopted the notation $(~)_{,t}\equiv d/dt$ and 
$(~)_{,tt}\equiv d^2/dt^2$ for simplicity of writing.
In this expression, the spatial curvature term  
reads, to leading order, as 
\beq
P_{\alpha }^{\beta }(t, x^{\gamma }) = \frac{1}{a^2(t)}
A_{\alpha }^{\beta }(x^{\gamma }) \, ,
\label{a5} 
\eeq 
where $A_{\alpha \beta }(x^{\gamma }) = {\xi }_{\beta \gamma}
A_{\alpha }^{\gamma }$ denotes the Ricci tensor corresponding 
to ${\xi }_{\alpha \beta }(x^{\gamma })$. 

The trace of (\ref{a4}) provides  the additional 
relation
\begin{align} 
2\left(a^3\right)_{,tt} 
+ \left(a^3{\eta}\right)_{,tt}\theta 
+ aA^\alpha_\alpha= %\nonumber\\
%&= 
\chi  \Bigg[ \frac{\epsilon}{3} \left( 3 + 4u^2 \right) 
+3\Lambda  \Bigg] 
2a^3\left( 1 + \frac{\eta \theta}{2} \right) \, .
\label{a6in} 
\end{align} 
Comparing (\ref{a2}) with the trace (\ref{a6in}), 
via their common term $(3+4u^2)\epsilon/3$, 
and estima\-ting the different orders of magnitude, 
we get the following equations: 
\bseq
\begin{align} 
\left(a^3\right)_{,tt} + 3a^2{a}_{,tt}- 
	4\chi a^3 \Lambda &= 0 
\label{a7a} \\
A_{\alpha\beta}&=0  
\label{a7b}\\
3\left(a^3\eta \right)_{,tt} + 3a^3{\eta }_{,tt} + 
	2\left(a^3\right)_{,t}{\eta}_{,t} + 
	9a^2\eta~{a}_{,tt} - 
	12 \chi a^3 \Lambda \eta &= 0 \, .
\label{a7c} 
\end{align}
\eseq
Since (\ref{a7b}) implies the vanishing of 
the three-dimensional Ricci tensor and this 
condition corresponds to the vanishing of 
the Riemann tensor too, then we can conclude that
the obtained Universe is flat 
to leading order, \idest 
\beq
\label{met}
\xi_{\alpha \beta}=\delta_{\alpha \beta} \, 
\qquad \Rightarrow \qquad j=1.
\eeq
Equation (\ref{a7a}) admits 
the expanding solution
\beq
a(t)=a_0 \exp\left( \frac{\sqrt{3\chi\Lambda}}{3}t\right)
\label{a9}
\eeq
$a_0$ being 
the initial value
of the scale factor
amplitude, taken at the instant $t=0$ 
when the de Sitter phase starts. 

Expression (\ref{a9}) for $a(t)$  
substituted in (\ref{a7c}) yields the 
differential equation for $\eta$ 
\beq
\ddot{\eta}+ \frac{4}{3}\sqrt{3\chi \Lambda}~\dot{\eta}=0 \, ,
\label{b2}
\eeq
whose only solution satisfying the
limit (\ref{tr}) reads as
\beq
\eta(t)=\eta_0 \exp\left(-\frac{4}{3}
\sqrt{3\chi\Lambda}~ t\right) 
\qquad \Rightarrow \qquad \eta = 
\eta_0 \left( \frac{a_0}{a}\right)^4 \, ,
\label{b3}
\eeq
and, of course, we require $\eta_0 \ll a_0$.

Equations (\ref{a1}) and 
(\ref{a2}), in view of the solutions (\ref{a9}) 
for $a(t)$ and (\ref{b3}) for $\eta(t)$, 
are matched for consistency by posing 
\bea 
u_{\alpha }(t, x^{\gamma }) = v_{\alpha }(x^{\gamma })
+ \mathcal{O}\left(\eta^2\right) \nonumber \\
(u_0)^2 = 1+ \mathcal{O}\left(\frac{1}{a^2}\right) 
\approx 1 \, , 
\label{b4}
\eea
and 
\beq
\epsilon = - \frac{4}{3} \Lambda\eta \theta \, ,
\label{b5}
\eeq
respectively, which implies $\theta < 0$ for each point 
of the allowed domain of the spatial 
coordinates.
The comparison between (\ref{a1}) 
and (\ref{b5}) leads to the explicit 
expression for $l(x^\gamma)$ in 
terms of $\theta$
\beq
\label{elle}
l(x^\gamma)=
\left(\frac{4}{3}\Lambda \eta_0{a_0}^4\right)^{3/4}
\left(-\theta\right)^{3/4} \, .
\eeq
Defining the auxiliary tensor with 
unit trace
${\Theta }_{\alpha \beta }(x^{\gamma })\equiv
{\theta }_{\alpha \beta }/\theta$,
the above analysis permits 
to obtain for it from (\ref{a4})
the expression
\beq
{\Theta }_{\alpha}^{ \beta }= 
\frac{\delta_{\alpha}^{\beta}}{3} \, .
\label{b6}
\eeq

By (\ref{xba}), the explicit 
form for $a$ expanded in $\eta$ 
yields the first-two
leading orders
of approximation for the scalar field
\begin{align}
\phi\left(t, x^\gamma\right)= 
{\cal{C}}\sqrt{\frac{t_r}{t_r-t}}
& \left(1- \frac{ 1}{4\sqrt{3\chi\Lambda}}
\frac{\eta}{t_r-t}\theta\right) \, , \\
& t_r =\frac{\sqrt{3\chi\Lambda}}{{\cal{C}}^22\lambda } \, ,
\nonumber
\label{b7}
\end{align}
where ${\cal{C}}$ is an integration constant; 
finally, equation (\ref{meeq3}) provides $v_{\alpha }$ 
in terms of $\theta$  
\beq
v_{\alpha }=-\frac{3}{4}\frac{1}{\sqrt{3\chi\Lambda}}
\partial_\alpha\ln\mid\theta\mid  \, .
\label{b8}
\eeq 

On the basis of (\ref{b6})-(\ref{b8}),
the hydrodynamic equations (\ref{montahydro})
%(\ref{or})-(\ref{qr}) 
reduce to an identity to leading order;
in fact such equations contain 
the energy density of
the ultra relativistic matter, known only
to first order (the higher one of the
Einstein equations). Therefore 
 it makes no sense to take into account 
 higher order contributions, coming from 
 those equations. 

As soon 
as $(t_r-t)$ is sufficiently large, the solution 
here constructed can be easily checked to be 
completely self-consistent to all the calculated 
orders of approximation in time and  contains one 
physically arbitrary function of the spatial coordinates, 
$\theta (x^{\gamma })$ which indeed, being a 
three-scalar, is not affected by spatial coordinate 
transformations.  \\
In particular, the terms quadratic in the spatial 
gradients of the scalar field are of order  
\beq
\label{ord}
\left(\partial_\alpha\phi\right) ^2 
\approx 
\mathcal{O}\left( \frac{\eta ^2}{a^2}
\frac{1}{\left(t_r-t
\right) ^3} \right)
\eeq
and therefore can be neglected with 
respect to all the inhomogeneous ones. \\
Such a solution fails when $t$ approaches
$t_r$ and ts validity requires 
the de Sitter phase to end (with the
fall of the scalar field in the true 
potential vacuum) when $t$ is still much 
smaller than $t_r$.%  (see below).

\subsection{Physical considerations}

The peculiar feature of the solution constructed lies 
in the independence of the function $\theta$ which, 
from a cosmological point of view, implies the 
existence of a quasi-isotropic inflationary solution 
in correspondence with an arbitrary spatial distribution 
of ultra-relativistic matter and of the scalar field. 

We get an inflationary picture from which the 
Universe outcomes with the appropriate standard features, 
but in presence of a suitable spectrum of {\it classical} 
perturbations as due to the small inhomogeneities which 
can be  modelled according to an Harrison--Zeldovich 
spectrum; in fact, expanding the function 
$\theta$ in Fourier series as 
\beq
\label{tet}
\theta(x^\gamma) = 
\frac{1}{(2\pi)^3}\int^{+\infty}_{-\infty}
{\tilde{\theta}\left(\vec{k}\right)
e^{i \vec{k}\cdot\vec{x}}d^3k} \, ,
\eeq
we can impose an 
Harrison--Zeldovich spectrum 
by requiring 
\beq
\label{hzi}
{\mid {\tilde{\theta}}\mid} ^2= 
\frac{Z}{\mid k\mid^{3/2}}\, , \qquad Z={\rm const.} \, .
\eeq

However, the following three points 
have to be taken into account to give 
a complete picture for our analysis:

%--------------------list---------
\newcounter{bean}
\begin{list}{(\roman{bean})}{\usecounter{bean}}%
\item 
        limiting (as usual) our attention 
        to leading order, the validity of 
        the slow-rolling regime is ensured
        by the natural conditions

        \begin{equation}
        \mathcal{O}\left(\sqrt{\chi \Lambda }(t - t_r)\right)\ll 1 \, ,
        \qquad
 \lambda \gg \mathcal{O}(\chi ^2\Lambda )
        \, , 
        \label{slr1}
        \end{equation}

        which respectively translate (\ref{infla2}) 
        and (\ref{infla1}); 
\item 
        denoting by $t_i$ and $t_f$ respectively 
        the beginning and
 the end of the de Sitter 
        phase, we should have $t_r\gg t_f$
        and the validity of our solution is 
        guaranteed if
        \newcounter{bean2}
        \begin{list}{\alph{bean2}}%{\usecounter{bean2}}%
        \item 
                (a) the flatness of the potential is preserved, 
                \idest
 $\lambda \phi ^4 \ll \Lambda $: such a 
                requirement coincides, 
as it should, with the 
                second of inequalities (\ref{slr1});
        \item 
                (b) given $\Delta$ as the width of the
                flat region of the potential,
 we require that the 
                de Sitter phase ends before
                $t$ becomes comparable with $t_r$, \idest

                \begin{equation}
                \phi (t_f) - \phi (t_i) \sim
                \sqrt{\frac{\sqrt{\chi \Lambda }}{\lambda }}
                \frac{t_f - t_i}{t_r^{3/2}} \sim \mathcal{O}(\Delta )
                \, , 
                \label{slr2}
                \end{equation}

                where we expanded the solution to first 
                order in
 $t_{i,f}/t_r$; via the usual position
                $(t_f - t_i) \sim 
                \mathcal{O}(10^2)/\sqrt{\chi \Lambda }$, 
                the relation (\ref{slr2}) becomes a constraint 
                for the
 integration constant $t_r$.
        \end{list}
\item 
In order to get an inflationary scenario, able 
to overcome the shortcomings present in the SCM, 
we need an exponential expansion sufficiently strong. 
For instance we have to require a region 
of space corresponding to a cosmological
horizon  $\mathcal{O}(10^{-24}cm)$ when the 
de Sitter phase starts to cover now all the
actual Hubble horizon $\mathcal{O}(10^{26}cm)$;
the redshift at the end of the de Sitter phase 
is  $z \sim \mathcal{O}(10^{24})$, then we should
require $a_f/a_i\sim e^{60}\sim \mathcal{O}(10^{26})$. \\
Let us estimate the density perturbations 
(inhomogeneities)
 at the (matter-radiation) 
decoupling age
 ($z\sim \mathcal{O}(10^{4})$) as
$\delta _{in}\sim \mathcal{O}(10^{-4})$; 
if we start by this
same value at the beginning  of inflation
($\delta ^i_{in}$),
 we arrive at the end with
$\delta ^f_{in} \sim (\eta _f/\eta _i)\delta ^i_{in} \sim
\mathcal{O}(10^{-100})$. 
Though these inhomogeneities
 increase as $z^2$ once 
they are at scale greater than the
 horizon, 
nevertheless they reach only
 $\mathcal{O}(10^{-60})$
at the decoupling age.  \\
This result provides support to the
idea that the spectrum of inhomogeneous perturbations cannot
have a classical origin in presence of an inflationary scenario.

\end{list}

%%%%%%%%%%%%%%%%%%%%%%%%%%%%ultima addenda%%%%%%%%%%%%%%%%%%%%%%%%%%%%%

In the considerations above developed,
we regard the ratio of the inhomogeneous terms 
$\epsilon_f$ and $\epsilon_i$ as the
quantity $\delta \rho $ %defined in Section 2 
and this assumption is
(roughly) correct: after the reheating,
the Universe is dominated by a homogeneous
(apart from the quantum fluctuations)
relativistic energy density $\rho _r$
to which is superimposed the relic $\epsilon _f$
after inflation; therefore we have
\begin{equation}
\delta \rho = \frac{\epsilon _f}{\rho _r} =
\frac{\epsilon _f}{\epsilon _i}
\frac{\epsilon _i}{\rho _r} =
\left(\frac{a_i}{a_f}\right)^4
\frac{\epsilon _i}{\rho _r} 
\, .
\label{axx}
\end{equation}
Hence our statement follows as soon as we observe
that the inhomogeneous relativistic energy density
before the inflation $\epsilon_i$ and the uniform 
one $\rho _r$, generated by the reheating process, 
differ by only some orders of magnitude.

%%%%%%%%%%%%%%%%%%%%%%%%%finis

\end{document}